\documentclass[aps,prl,twocolumn,showpacs,superscriptaddress,groupedaddress]{revtex4-1}
\pdfoutput=1
\usepackage{pgf}
\usepackage{tikz}
\usetikzlibrary{calc,patterns,decorations.pathmorphing,decorations.markings}
\usetikzlibrary{arrows}
\usetikzlibrary{arrows.meta}

\usepackage[caption=false]{subfig}
\usepackage{graphicx} 
\usepackage{mathrsfs}
\usepackage{amsfonts}
\usepackage{amsmath}
\usepackage[noabbrev]{cleveref}
\usepackage[toc,page]{appendix}
\usepackage{import}
\usepackage[nice]{units}
\usepackage{float}
\def\s{\sigma}  \def\a{\alpha} \def\b{\beta}
 \def\d{\delta} \def\l{\lambda}

\def\e{\varepsilon}

\def\gbar{\bar{g}}

\def\gbaro{\bar{g}^0}

\def\nn{\nonumber \\}

\begin{document}
\title{Delayed instabilities in viscoelastic solids through a metric description}
\date{\today}
\author{Erez Y. Urbach}
\author{Efi Efrati} 
\email{efi.efrati@weizmann.ac.il} 
\affiliation{Department of
Physics of Complex Systems, Weizmann Institute of Science, Rehovot
76100, Israel}

\begin{abstract} 
%\textbf{version 1}\\
%Many manmade and naturally occurring biological elastomers form viscoelastic solids. The variety of applications and the extreme mechanical response such elastomeric materials offer drive the need for a better, more intuitive and quantitative understanding of the mechanical response of such continua. In particular, we presently have a very limited theoretical understanding and quantitative predictability of stability and delayed stability loss in viscoelastic solid structures. 
%In this work we put forward a metric description of viscoelasticity in which the continua is characterized by temporally evolving reference lengths with respect to which elastic strains are measured. Formulating the three dimensional theory using metric tensors we are able to predict which structures will exhibit delayed stability loss due to viscoelastic flow. We also quantitatively describe the viscoelastic relaxation in free standing structures including cases where the relaxation leads to no apparent motion. We conclude by demonstrating the power of the theory by elucidating the subtle mechanism of delayed stability loss in elastomer shells and show quantitative agreement with experimental results. \\
%\textbf{version 2}\\
While determining the stability of an unconstrained elastic structure is a straightforward task, this is not the case for viscoelastic structures. Seemingly elastically stable conformations of viscoelastic structures may gradually creep until stability is lost, and conversely, creeping does not necessarily imply that a structure will eventually become unstable. Understanding instabilities in viscoelastic structures requires a more intuitive description of viscoelasticity to allow analytical results and quantitative predictions.
In this work we put forward a metric description of viscoelasticity in which the continua is characterized by temporally evolving reference lengths with respect to which elastic strains are measured. Formulating the three dimensional theory using metric tensors we are able to predict which structures will exhibit delayed instability due to viscoelastic flow. We also quantitatively describe the viscoelastic relaxation in free standing structures including cases where the relaxation leads to no apparent motion. We demonstrate these results and the power of the metric approach by elucidating the subtle mechanism of delayed instability in elastomer shells showing quantitative agreement with experimental measurements.\end{abstract}
\maketitle
%
% 
%Stationary states and delayed instability in viscoelastic solids: A metric approach
%\\
%Viscoelastic solids are capable of displaying transiently stable states, that creep visco-elastically until 
%lose their 
%
%
%

\paragraph{Introduction}

The snapping of the Venus fly-trap leaf, one of the fastest motions in plant kingdom, is preceded by a relatively slow creeping motion~\cite{FSDM05}. A similar creep is observed prior to the snap through of thin elastomeric shells known as jumping poppers \cite{PMVH14}. While the snap through itself lasts only a fraction of a second~\cite{GMV16,PMVH14}, the slow creeping motion, during which the shells seems to be elastically stable, may last orders of magnitude longer. On a much larger scale, similar behavior is in some cases attributed to earth's crust prior to an earthquake \cite{FL01,RSA94}. In all of these systems the viscoelastic flow in the material serves as the instigator for the instability that releases the internally stored elastic energy. While in most cases we are able explicitly write down the equations governing the viscoelastic behavior, the process of delayed viscoelastic instability remains poorly understood.

Viscoelastic solids display reversible elastic behavior when loads at a fast rate. However, they also show a slow creeping behavior under a constant load, and exhibit stress relaxation when held at constant displacement. 
Commonly, such solids are modeled by a constitutive law relating the stress rate to the stress, strain and strain rate, or equivalently by expressing the stress as a function of the history of strain rate and a material dependent memory kernel~\cite{RC12}. While these approaches capture the material response well, and yield accurate results through simulations of viscoelastic structures~\cite{MS10,BSP12}, they are rarely explicitly solvable and provide little intuition regarding the state of the viscoelastic material and in particular have little in the way of clarifying the governing processes in viscoelastic instabilities. Recent attempts to address viscoelastic instabilities by 
modeling the viscoelastic response as an elastic media with temporally evolving stiffness~\cite{BSP12,MS10,KLK10} show only a qualitative agreement with experiment, and have limited applicability as in particular they fails to capture creeping at zero load.

\begin{figure}[H]
	\resizebox{\linewidth}{!}{\includegraphics{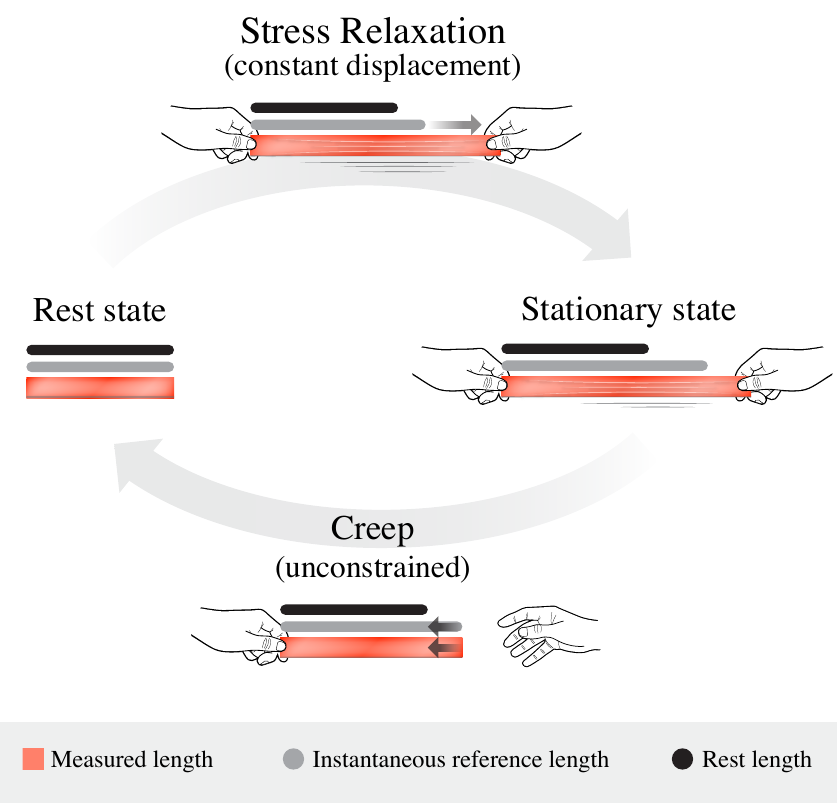}}

	\caption{{\em The viscoelastic  reference length evolution.} At the resting state all three length measures on the body, its measured length $g$ (marked red), its instantaneous reference length $\gbar$ (marked grey), and its rest reference length $\gbaro$ (marked black) are all equal.   
	When subjected to a constant displacement extension, the instantaneous reference length evolves away from the rest length and towards the presently assumed length, thus resulting in stress relaxation. It asymptotically approaches the stationary state $\gbar^\text{stat}= \b g+ (1-\b)\gbaro$, in which the initial stress is reduced by a factor of $1-\b$. 
When released the unconstrained system immediately adopts its favored instantaneous reference length, which in turn gradually creeps towards the rest lengths.}
	\label{fig:1d_illustration}
\end{figure}

In this works we quantitatively describe the notion viscoelastic instability through a metric description. In this metric approach we describe the materials' behavior as a fast elastic response with respect to temporally evolving rest lengths that change due to the slow viscoelastic flow. In this framework the microscopic response of material that leads to stress relaxation is interpreted as the evolution of the rest lengths of the system towards the lengths assumed in its present state (see FIG.~\ref{fig:1d_illustration}).
The rest lengths in the system serve as a state variable accounting for the slow and dissipative evolution in the material.   In particular it allows us to analyze the notion of stability in viscoelastic materials and discover a new stationarity property; if a viscoelastic system starting at rest is brought abruptly to another locally stable state, then despite the continuous evolution of the internal stresses in the structure, it will display no motion. This result in particular elucidates the subtle nature of delayed instabilities in such systems, and allows a quantitative understanding of them as we demonstrate through experiment.

\paragraph{Linear viscoelastic continua}
All linear viscoelastic material constitutive relations, and in particular the relations for every spring dashpot model, can be expressed through a convolution of the strain rate with a stress relaxation function, $\s(t) = \int_{-\infty}^t G(t-s) \dot \e(s) ds$~\cite{AB09,RC12}. This relation can be  brought  into the form~\cite{CN61}
\begin{equation}
	\s(t) = C \left(\e(t) + \beta\int_{-\infty}^{t} \dot \phi(t-s) \e(s)ds \right). \label{eq:linear_viscoelasticity} 
\end{equation}
Here $C$ is the material stiffness, $0<\b<1$ is a dimensionless constant of the material, and $\phi(s)$ is the \emph{normalized memory kernel} that satisfies
%
%The memory of events should decrease over time and thus we take
 $\phi(0)=1$, $\phi(\infty)=0$ and as we expect the memory to only decay in time also satisfies $\dot \phi(s) <0$.
For a general three dimensional isotropic and incompressible material the tensorial generalization of  \eqref{eq:linear_viscoelasticity} reads~\cite{UE18b}:
%
%To describe a general three dimensional media, $C$ and $\phi(s)$ need to be regarded as tensors. For incompressible materials~\cite{UE18b}, the general constitutive relation between stress and strain 
\begin{equation}
	S^{ij}(t) = C^{ijkl} \left( \e_{kl}(t) + \b \int_{-\infty}^{t} \dot \phi(t-s) \e_{kl}(s)ds \right) . \label{eq:tensor_linear_viscoelasticity}
\end{equation}
Here $C^{ijkl}$ is the isotropic stiffness tensor
\footnote{Assuming the material's response is isotropic and homogeneous near its rest stationary state, the stiffness tensor take the form $C^{ijkl} = \l \ \gbar_0^{ij} \gbar_0^{kl} + 2\mu \  \gbar_0^{ik} \gbar_0^{jl}$, where $\gbar_{0}^{ij}$ is the inverse rest reference metric.
We note that one could assume the material to be isotropic and uniform with respect to other metrics, for example with respect to $g_{ij}$ (i.e. in the lab frame). Such theories will differ from the one presented here only in higher orders of the strain. The choice we make here is the most natural generalization for temporally evolving metrics, and is computationally favorable.}.
$S^{ij}(t)$ denote the second Piola-Kirchhoff stress tensor, which through the equation above depends on the full history of the 
the strain tensor 
\begin{equation}
	\e_{ij}(t) = \tfrac 1 2 \left(g_{ij}(t)-\gbaro_{ij} \right). \label{eq:strain_relation}
\end{equation}
The metric $g_{ij}$ measures lengths in the material and 
 uniquely describes the configuration of the elastic body.  We have also introduced 
the \emph{rest reference metric}, $\gbaro_{ij}$, the metric at which the system is locally stress-free and stationary. 
That is, if the elastic solid was to be cut to infinitesimal pieces and each of the pieces allowed to freely relax indefinitely, then the lengths in each of the pieces as given by the metric  would approach the lengths given by the rest reference metric, $\gbaro_{ij}$. For a more detailed description of  covariant elasticity using metric tensors the reader is referred to \cite{UE18b, SE10}.

Eq.~\eqref{eq:tensor_linear_viscoelasticity} predicts that instantaneous incremental deformations $\Delta g_{ij}$ lead to 
linear stress increments $\Delta S^{ij} = C^{ijkl} \tfrac{1}{2} \Delta g_{kl}$ as one would predict for a purely elastic response. Inspired by this instantaneous elastic-like response we define the time-dependent \emph{reference metric} $\gbar_{ij}$ of the body such as to satisfy
\begin{equation}
	S^{ij}(t) = C^{ijkl}\tfrac{1}{2} \left(g_{kl}(t)- \gbar_{kl}(t) \right). \label{eq:stress_relation}
\end{equation}
The temporal evolution of the reference metric can be deduced from \cref{eq:tensor_linear_viscoelasticity,eq:strain_relation,,eq:stress_relation}, and reads
\begin{equation}
	\gbar_{ij}(t) = (1-\b)\ \gbaro_{ij} - \b \int_{-\infty}^{t} \dot \phi(t-s)\ g_{ij}(s)\ ds. \label{eq:gbar_evolution_rule}
\end{equation}
It is important to stress that equations \eqref{eq:stress_relation} and \eqref{eq:gbar_evolution_rule} are completely equivalent to Eq.~\eqref{eq:tensor_linear_viscoelasticity}. However the notion of a reference metric allows a more intuitive understanding of viscoelastic behavior. 
Note that $\gbar_{ij}(t)$ remains unchanged by instantaneous variations of $g_{ij}$. We may thus consider it as the slow state variable describing the viscoelastic evolution of the material. At each moment in time we may consider the system as an elastic system with respect to the reference metric $\gbar_{ij}$~\cite{ESK09,ESK13}.
The dimensionless factor $\beta$ can be shown by Eq.~\eqref{eq:tensor_linear_viscoelasticity} to quantify the fraction of stress asymptotically relaxed in a constant displacement experiment starting from rest. $\b$ is thus a measure of the degree of viscoelasticity in the system. For a prescribed configuration given by the metric $g_{ij}$ the reference metric $\gbar_{ij}$  will asymptotically approach the stationary solution of  Eq.~\eqref{eq:gbar_evolution_rule}:

\begin{equation}
	\gbar^\text{stat}_{ij} = \b g_{ij} + (1-\b)\gbaro_{ij}. \label{eq:gbar_stationary_condition}
\end{equation}
Therefore, $\b$ also measures the material's propensity to change its reference metric toward the realized metric in its present configuration.
$\b=0$ describes a purely elastic material, where the reference metric remains $\gbar_{ij}=\gbaro_{ij}$ for all times~\footnote{The body is assumed to start with $\gbar_{ij} = \gbaro_{ij}$. Alternatively, $\gbar_{ij}$ relaxes to $\gbaro_{ij}$ independently of the behavior of the body $g_{ij}$, thus after long enough relaxation we can take $\gbar_{ij} = \gbaro_{ij}$.}. $\b=1$ describes a viscoelastic fluid where the rest reference metric $\gbaro_{ij}$ has no meaning and the material approaches arbitrarily close to $g_{ij}$, diminishing the stress to zero in relaxation.

\paragraph{The quasi static approximation}
Viscoelastic systems are dissipative, thus the notion of an elastic free energy is ill defined. Nonetheless, the virtual work of displacements performed over a short period $\Delta t \rightarrow 0 $, coincides with the instantaneous elastic energy functional~\cite{ESK09}
\begin{equation}
	E\left[\e^\text{el}_{ij}\right]=\int \frac{1}{2}C^{ijkl} \e^\text{el}_{ij} \e^\text{el}_{kl}\ \sqrt{|\gbaro|}\ d^3x,
	\label{eq:energy_functional}
\end{equation}
where the elastic strain is $\e^\text{el}_{ij} = \tfrac{1}{2}\left(g_{ij}-\gbar_{ij}\right)$. Typically the elastic response time scale in elastomers is much smaller than the viscoelastic relaxation time. In such cases we can eliminate inertia from the system and approximate the motion of the material as quasi static evolution between elastic equilibrium states. That is, the configuration at every instance in time, given by the metric $g_{ij}(t)$, minimizes the instantaneous elastic energy functional \eqref{eq:energy_functional}. If we can characterize % I want to say locally in the manifold of metrics...
 the possible Euclidean metrics by finite set of variables $\l_\a$, $g_{ij}(x) = g_{ij} [ \l_\a ] (x)$ then the minimization condition is for all $\a$, $\frac{\delta E}{\delta \l_\a}=0$ or
\begin{equation}
	\frac{1}{2}\int \frac{\delta g_{ij}}{\delta \l_\a} S^{ij} \ \sqrt{|\gbaro|}\ d^3x =0. \label{eq:minimaztion_condition}
\end{equation}
For this extremal point to be a minima, the Hessian, $\frac{\d^2E}{\d \l_\a \d \l_\b}$, needs to be positive definite.
Condition \eqref{eq:minimaztion_condition} gives $g_{ij}(t)$ as function of $\gbar_{ij}(t)$. $\gbar_{ij}(t)$ on the other hand evolve in time according to Eq.~\eqref{eq:gbar_evolution_rule}. 
The quasi static coupled evolution of $g_{ij}$ and $\gbar_{ij}$ is simpler to analyze both numerically and theoretically compared with the full dynamics.

As seen from Eq.~\eqref{eq:gbar_stationary_condition}, the long time stability of a body depends only on $\b$, and not on the specific memory kernel. It is thus instructive to choose the model of standard linear solid (SLS), which is the simplest model capable of exhibiting creeping motion, a finite stress relaxation characterized by $0<\b<1$ and possess only a single time scale $\tau$ \cite{AB09,RC12}.  
Its memory kernel is given by $\phi(s)=e^{-\nicefrac{s}{\tau}}$. The constitutive laws of the SLS model relating the stress and stain can be recast as an evolution law for the reference metric 
\begin{align}
	\dot \gbar_{ij}(t) =& -\frac{1}{\tau}\left(\b \left(\gbar_{ij}(t) - g_{ij}(t) \right) +(1-\b) \left(\gbar_{ij}(t) - \gbaro_{ij} \right)\right) \nn
	=&-\frac{1}{\tau}\left( \gbar_{ij}(t)-\gbar^\text{stat}_{ij} (t)\right).
	\label{eq:gbar_SLS_DE}
\end{align}
This equation is identical to Eq.~\eqref{eq:gbar_evolution_rule} under the SLS memory kernel, cast in a more intuitive differential equation.

\paragraph{Metastable configuration-stationarity}
A given reference metric $\gbar_{ij}$ can yield multiple elastically stable configurations in the instantaneous elastic energy functional \eqref{eq:energy_functional}. As the reference metric evolves viscoelastically according to Eq.~\eqref{eq:gbar_evolution_rule} it could acquire new stable configurations, merge existing stable points or cause stable elastic configurations to lose their stability. The latter phenomenon, termed \emph{delayed instability}, manifests as a slow viscoelastic evolution followed by a rapid snap to a near stable configuration. This phenomenon was also termed temporary bistability~\cite{MS10}, pseudo bistability~\cite{BSP12} or creep buckling~\cite{NH56,BH78} and forms the main difficulty in predicting the stability of visco-elastic structures. Here we show that the question of stability of the viscoelastic system reduces to examining the purely elastic stability of the rest system where $\gbar_{ij}=\gbaro_{ij}$.   

\begin{figure}[t]
	\resizebox{\linewidth}{!}{\newcommand*{\pgfmathsetnewmacro}[2]{%
    \newcommand*{#1}{}% Error if already defined
    \pgfmathsetmacro{#1}{#2}%
}

\begin{tikzpicture}[scale=2]
	\pgfmathsetnewmacro{\len}{3}
	\pgfmathsetnewmacro{\rad}{3}
	\pgfmathsetnewmacro{\ang}{25}
	\pgfmathsetnewmacro{\gang}{\ang*.25}
	\pgfmathsetnewmacro{\circlerad}{.05}
	\pgfmathsetnewmacro{\m}{tan(\ang/2)}
	\pgfmathsetnewmacro{\xo}{\rad*cos(\gang)}
	\pgfmathsetnewmacro{\yo}{{\rad*sin(\gang)}}

	\pgfmathsetnewmacro{\rado}{2}
	\pgfmathsetnewmacro{\ango}{20}
	\pgfmathsetnewmacro{\xoo}{\xo+\len+\rado*cos(\gang)+\rado*cos(180+(\gang-\ango)*.5)}
	\pgfmathsetnewmacro{\yoo}{\yo+\m*\len+\rado*sin(\gang)+\rado*sin(180+(\gang-\ango)*.5)}

	% Arcs
	\draw (\xo+.6*\len,\yo+\m*.6*\len) node(gstat){} circle[radius=\circlerad];

	\draw[line width=2.5] (\rad,0) arc(0:\ang:\rad)
	(\xoo, \yoo) arc(180+(\gang-\ango)*.5:180+(\gang+\ango)*.5:\rado);

	% Cirvles
	\draw[fill=black] (\xo,\yo) node(g){} circle[radius=\circlerad*1.3];
	\draw[fill=red,color=red] (\xo+\len,\yo+\m*\len) node(gbar0){} 
			circle[radius=\circlerad*.8];
	\draw[color=black!50,fill=black!50] (\xo+.9*\len,\yo+\m*.9*\len) node(gbar){} 
							(gbar.center) circle[radius=\circlerad];

	\draw[dotted] (g) node[below=6pt,right=2pt]{$g$} 
	-- (gstat) node[right=4pt,below=1pt]
	{$\bar g^{\text{stat}}$}
	-- (gbar) -- (gbar0) node[right=10pt,below=-10pt]{$\bar g^0$};
	\draw (gbar.center) node[above=2pt]{$\bar g(t)$};
	\draw[-latex,line width=2,color=black!50] (gbar.center) -- (\xo+.8*\len,\yo+\m*.8*\len);
	
	\draw[anchor =south west] ({\xo*.97},{\yo*2.4}) node[text width=1.5cm]
	{\scriptsize Admissible metrics};
\end{tikzpicture}}
	\caption{A schematic representation of the metrics collinearity. The  minimization of the metric $g$ (marked by a full black circle) is constrained and performed with respect to the subset of metrics that correspond to realizable configurations (thick black line). Such metrics are in particular orientation preserving and Euclidean. Given a reference metric $\gbar$ (marked by a full gray circle) the realized metric will correspond to the closest point from the set of admissible metrics to $\gbar$ according to the distance function given by the instantaneous elastic energy \eqref{eq:energy_functional}. Starting from rest, $\gbar$ evolves from $\gbaro$ towards the $g$, which remains the closest admissible metric to $\gbar$ due to collinearity of the three metrics. As $g$ remains stationary, the evolution of $\gbar$ will preserve the collinearity, approaching asymptotically $\gbar^{\text{stat}}$, which is also collinear. We stress that throughout this evolution $g$ remains unchanged, thus no variation of the configuration will be observed despite the stress relaxation.
		}\label{fig:metric_space}
\end{figure}
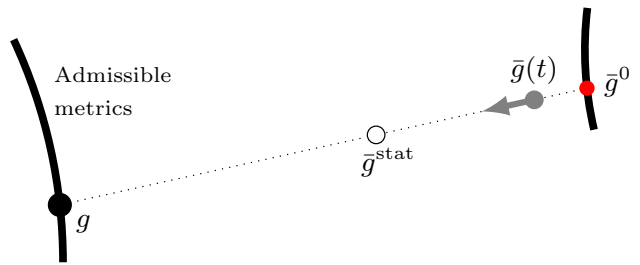

Starting from rest, and deforming fast into a local elastic stable configuration will result in a stationary configuration, i.e. a configuration that will display no temporal variation despite the continuous evolution of the  reference metric and the relaxation of the corresponding stress~\footnote{For the case of elasto-plasticity, a similar idea was first proved but never published by H. Aharony. This type of behavior was also conjecture~\cite{BH78}.}. To show this claim it suffices to assume that $g_{ij}$ in Eq.~\eqref{eq:gbar_evolution_rule} is constant in time to obtain $\gbar_{ij}(t)$ and show that the constant $g_{ij}$ satisfies the equilibrium equation \eqref{eq:minimaztion_condition} with respect to the obtained  $\gbar_{ij}(t)$.
Conversely, if a system converges to a fixed stable state (where the configuration as well as the reference metric do not evolve), the corresponding  configuration must also be extremal with respect to the rest reference metric. This again could be shown by substituting Eq.~\eqref{eq:gbar_stationary_condition} into Eq.~\eqref{eq:minimaztion_condition}.
We note that these results are not limited to a particular memory kernel. Intuitively, both claims are due to collinearity of $g_{ij},\gbaro_{ij}$ and $\gbar_{ij}$ under the appropriate initial conditions, see FIG.~\ref{fig:metric_space}.
For explicit proof see \cite{UE18b}. 

\begin{figure}[t]
	\centering	
	\begin{minipage}{\linewidth}
		\begin{tikzpicture}
		    \node[anchor=south west,inner sep=0] at (0,0) {\includegraphics[width=\linewidth]{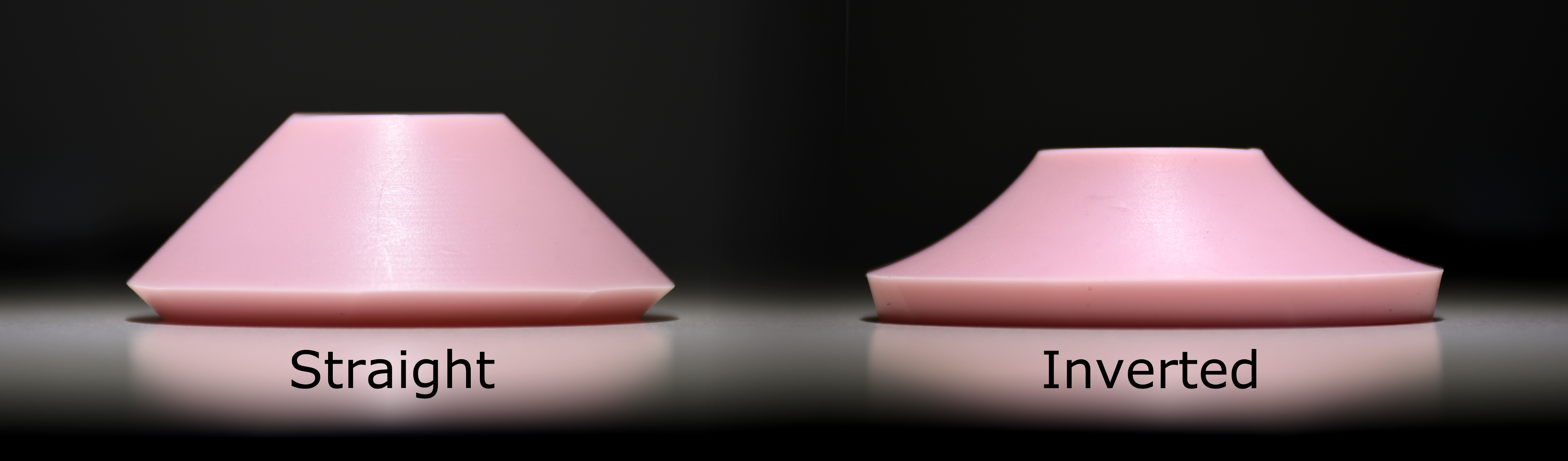}};
		    \node[text=white] at (.3,2.25) {(a)};
		\end{tikzpicture}
		 \hfill\null
		\begin{tikzpicture}
		    \node[anchor=south west,inner sep=0] at (0,0) {\resizebox{\linewidth}{!}{\input{./results_beta_0.1.pgf}}};
		    \node[text=black] at (.3,4.6) {(b)};
		\end{tikzpicture}
		 \hfill\null
		\begin{tikzpicture}
		    \node[anchor=south west,inner sep=0] at (0,0) {\resizebox{\linewidth}{!}{\import{}{flip_time_vs_thickness.pgf}}};
		    \node[text=black] at (.3,3.9) {(c)};
		\end{tikzpicture}
	\end{minipage}
	\caption{\label{fig:hopper_popper}
	(a)~Straight and inverted conical poppers.
	(b)~The two axes span the dimensionless geometrical properties of the truncated conical poppers. The background colors represent the theoretically predicted regions of each of the phases. Each marker corresponds to a different popper, different shaped (and colored) markers indicate the different phases observed in experiment.
	(c)~Numerically calculated flipping time as function of the normalized thickness of the conical popper for immediate release and long holding time. The different poppers were simulated by varying their thicknesses and constant radii $r_\text{min}=10[\text{mm}],r_{max}=25[\text{mm}]$. The material properties taken were $\b = 0.1$, $\tau=0.1[\text{sec}]$, Young's modulus $E=2.5[\text{MPa}]$ and Poisson ratio $\nu=0.47$. A similar phase plot presented in \cite{BSP12}.
	}
\end{figure}

\paragraph{Acquired elastic stability and delayed instability} 
When an appropriately cut elastic cone is flipped inside out and laid on the table (FIG.~\ref{fig:hopper_popper}a) it creeps for a few seconds and snaps back to its original shape. This phenomena of delayed instability seems to contradict the metastable configuration stationarity described above. However, when examined carefully this stationarity property is not only reconciled with observations but to a large degree explains them quantitatively. A particular corollary of the configuration stationarity property is that instability could be observed only for initially unstable configurations. Had the system displayed stability at its initial state one can show that the state remains a local minimum of the elastic energy.  
Thus delayed instability can only be observed for states whose stability was acquired by the viscoelastic evolution \cite{MS10,BSP12}. One can show that acquired stability in configurations that were not extremal initially (e.g. saddle points) cannot persist indefinitely and  must be eventually lost.

\paragraph{Experimental results}

Our theory predicts that when examining viscoelastic structures, initially stable configurations will show no temporal evolution of the configuration and in particular always keep their stability. When held fixed in a non-extremal configuration stress relaxation could lead to new acquired stable configurations, whose stability is lost in time. These results hold for all incompressible linear viscoelastic bodies regardless of the form of the stress relaxation kernel. We next come to test these predictions by examining experimentally the response of silicone-rubber poppers.

We cast silicon rubber poppers  in the geometry of truncated-conical shells (FIG.~\ref{fig:hopper_popper}a). The conical shells have an apex angle of $45^\circ$, inner and outer radius radii $r_\text{min},r_{max}$ and thickness $h$. Sufficiently thin poppers show bistability with the inverted shape close to a mirror image of the undeformed state. As the thickness is increased this bistability is broken and  when brought from rest to the inverted state the popper immediately snaps back. If the thickness is large enough this instability will persist regardless of how long we hold the conical shell in its inverted state. For intermediate values of the thickness we expect to observe unstable states that could acquire stability if held long enough in their inverted state. This acquired stability is expected to be lost in a finite time. These three phases are plotted in FIG.~\ref{fig:hopper_popper}c. 

%To test the notion of the metastable configuration-stationarity we produce two popper of very close geometries one stable when inverted from equilibrium and the other unstable. We invert the stable popper fast and examine its motion using a camera. The unstable popper is held inverted for 30 seconds to acquire stability and its evolution is also followed. While the unstable popper shows a visible significant creep prior to snapping, no discernible motion was observed for the stable cone. See supplementary movie. 

The conical shell popper geometry is completely determined by two dimensionless variables $r_{min}/h$ and $r_{max}/r_{min}$. The boundaries between the different stability phases of FIG.~\ref{fig:hopper_popper}c depend on these geometric parameters as well as on the material constant $\beta$. We produced $\sim 50$ different conical popper of different geometries and tested their phases. All popper were produced from the same material and share the same $\b$. Starting from rest $g_{ij}=\gbar_{ij}=\gbaro_{ij}$ the poppers were abruptly brought to the inverted shape, and held for time $t_\text{hold}\sim1$ min. The poppers were then released, and the time to snap though $t_\text{flip}$ was measured. Determining the phases boundaries requires knowledge of $\beta$, which we measured through a stress relaxation experiment yielding  $\b = 0.0928 \pm 0.0077$. Using this value we calculated numerically the expected position of the phase boundaries. FIG.~\ref{fig:hopper_popper}b shows the experimentally obtained phases superimposed on the phases predicted from the theory showing very good agreement.

\paragraph{Discussion} 
The metric description of viscoelasticity presented here is reminiscent of the additive decomposition of strains used to study elasto-plasticity as for example is done in~\cite{EL69},
\begin{equation}
 	\e_{ij} =\e^\text{el}_{ij}+\e^\text{ve}_{ij}=\tfrac{1}{2}\left(g_{ij}-\gbar_{ij}\right)+\tfrac{1}{2}\left(\gbar_{ij}-\gbaro_{ij}\right),
\end{equation} 
where $\e^\text{el}_{ij}$ and $\e^\text{ve}_{ij}$ are the elastic and viscoelastic strains respectively. One can generalize the theory to account for plasticity or growth within the material by allowing 
$\gbaro_{ij}(t)$ to vary in time resulting in a covariant visco-elasto-plastic theory. The configurational response in such growth processes shows the evolution of $\gbar_{ij}$ rather than directly that of $\gbaro_{ij}$. Thus the viscoelastic response may cause a simple growth rule to appear complex, or cause an abrupt and localized growth events to seem diffuse and gradual. 

The covariant description of viscoelasticity employing the notion of the reference metric $\gbar_{ij}$  applies for all linear-viscoelastic materials. The result of metastable configuration stationarity was, however, obtained under the restricting assumptions that the Poisson ratio associated with the elastic and viscoelastic response are similar, and that
the system is brought instantaneously from its rest state to the metastable state. The former assumption may seem plausible for incompressible elastomers and gels, yet the later is expected to be violated by every physical experiment. While one could not prove full stationarity if these assumptions are lifted, the material's response is continuous in the deviations from the idealized conditions and small deviations will result in very little motion, see \cite{UE18b} for more details. In particular in the experiment we performed no discernible motion was observed for metastable state. 

The theory proves particularly powerful when applied to describe the delayed instability in elastomeric thin shells. We have focused on stability classification in the present work, yet predict to be able to quantitatively describe the non-trivial dependence jumping time on the holding time in the inverted state, and in particular its divergence near the boundary between the unstable regimes and stable regime in FIG.~\ref{fig:hopper_popper}c. 

The coupling of discrete singular fast deformations to slow creeping evolution was recently suggested to underlie non-monotonous and history dependent evolution in athermal systems, allowing them to display glassy behavior without the need to invoke the notion of effective temperature~\cite{LGAR17}. The analysis below ties this creeping evolution and elastic instability, to the notion of frustration, and the existence of a realizable stress free state, constraining the possible response such media can yield. Thus, it may provide insight not only to viscoelastic dynamics and instabilities of macroscopic systems, but may also apply to the meso-scale evolution of creeping athermal systems~\cite{KWWN02}.

%\begin{acknowledgments}
%The authors would like to acknowledge D. Biron, G. Cohen, A. Grosberg, S.M. Rubinstein, E. Sharon, Y. Bar-Sinai and D. Vella for helpful discussions, as well as the Aspen Center for Physics, which is supported by National Science Foundation grant PHY-1607761, for hosting many of these discussions. 
%%E.E. is supported by 
%\end{acknowledgments}

\bibliography{ourBIB}

\end{document}